\documentclass[journal,comsoc]{IEEEtran}
\usepackage[T1]{fontenc}

\ifCLASSINFOpdf
  \usepackage[pdftex]{graphicx}
  \graphicspath{{../pdf/}{../jpeg/}}
  \DeclareGraphicsExtensions{.pdf,.jpeg,.png}
\else
  \usepackage[dvips]{graphicx}
  \graphicspath{{../eps/}}
  \DeclareGraphicsExtensions{.eps}
\fi

\usepackage{amsmath}

\usepackage{bm}

\newtheorem{theorem}{Theorem}

\newtheorem{proposition}[theorem]{Proposition}
\usepackage{algorithm}  
\usepackage{algorithmicx}  
\usepackage{algpseudocode}

\usepackage{url}
\usepackage{stfloats}
\usepackage{cite}

\begin{document}
\bstctlcite{BSTcontrol}

\title{Reconfigurable Intelligent Surfaces for Energy Efficiency in Full-duplex Communication System}
\author{
	\IEEEauthorblockN{Yiru Wang, \IEEEmembership{Graduate Student Member,~IEEE,} Pengxin Guan, Hongkang Yu and Yuping Zhao}
	\thanks{This work was supported by Pengcheng National Laboratory No.PCL2021A04 of China. \emph{(Corresponding author: Yuping Zhao.)}}
	\thanks{Yiru Wang, Pengxin Guan and Yuping Zhao are with the School of Electronics Engineering and Computer Science, Peking University, Beijing 100080, China. (email: yiruwang@stu.pku.edu.cn; guanpengxin@pku.edu.cn; yuping.zhao@pku.edu.cn).}
	\thanks{Hongkang Yu is with Wireless Product Research and Development Institute, ZTE Corporation, Shenzhen 518057, China. (email: yu.hongkang@zte.com.cn).}
	

}

\markboth{Journal of \LaTeX\ Class Files,~Vol.~14, No.~8, August~2015}%
{Shell \MakeLowercase{\textit{et al.}}: Bare Demo of IEEEtran.cls for IEEE Communications Society Journals}

\maketitle

\begin{abstract}
In this letter, we study the reconfigurable intelligent surfaces (RIS) aided full-duplex (FD) communication system. By jointly designing the active beamforming of two multi-antenna sources and passive beamforming of RIS, we aim to maximize the energy efficiency of the system, where extra self-interference cancellation power consumption in FD system is also considered. We divide the optimization problem into active and passive beamforming design subproblems, and adopt the alternative optimization framework to solve them iteratively. Dinkelbach’s method is used to tackle the fractional objective function in active beamforming problem. Penalty method and successive convex approximation are exploited for passive beamforming design. Simulation results show the energy efficiency of our scheme outperforms other benchmarks.
\end{abstract}


%
\IEEEpeerreviewmaketitle

\section{Introduction}

\IEEEPARstart{W}{ith} the vision of 6-th Generation, the energy efficiency (EE) has been widely used as an important metric in green communications \cite{Green}. To improve the EE performance, a low-energy equipment named reconfigurable intelligent surface (RIS) has gained in popularity, which can dynamically change the wireless channels by adjusting the phase shifts \cite{RIS2}. A few literature have studied the EE maximization problem in RIS-aided system \cite{PowerRIS,dtd}. The authors of \cite{PowerRIS} investigated the EE optimization of RIS-aided downlink multi-user communication from a multi-antenna base station. The EE design for a RIS-aided device-to-device half-duplex (HD) communication network was proposed in \cite{dtd}.

Theoretically, full-duplex (FD) technology can double the spectral efficiency by transmitting and receiving signals over the same time-frequency dimension, thus it further improves the communication performance compared to HD mode \cite{intro1,intro2}.  However, FD systems would suffer from strong self-interference (SI) signal in practice. A number of SI cancellation (SIC) methods can suppress the SI power to the noise floor \cite{FDSI}, but would also cause extra power consumption \cite{SIC}. The optimization of the system EE for a RIS assisted FD communication system has been studied in \cite{EEFD}, where each device transmit and receive signals with separative single antenna. However, the additional power consumption induced by SIC in FD system has not been considered. Futhermore, how to jointly design active beamforming and passive beamforming to  improve EE in a RIS-aided multi-antenna FD system is still a challenge. 

This work studies the EE maximization problem of multi-antenna point-to-point FD communication system. The main contributions are summarized as follows:

\begin{itemize}
	\item We study the EE maximization problem in RIS-aided FD system, subject to the minimum data rate demands and maximum transmit power constraints, together with the unit-modulus constraint at the RIS. The extra power consumption induced by SIC is also considered.
	\item We decouple the non-convex problem into active and passive beamforming subproblems and use the alternative optimization (AO) framework to solve them iteratively. We use Dinkelbach’s method to tackle the fractional objective function in active beamforming problem. Penalty method and successive convex approximation (SCA) are exploited for passive beamforming design.
    \item Simulation results show the EE performance of RIS aided FD system is superior than other benchmarks.
\end{itemize}

\begin{figure}
	\centering
	\includegraphics[width=2.0in]{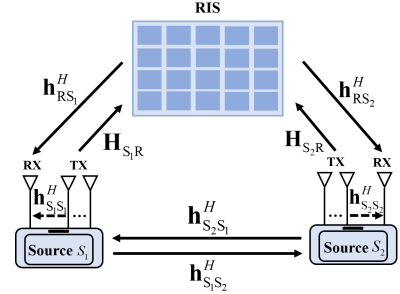}
	\caption{A RIS assisted multi-antenna FD communication system.}
\end{figure}

\emph{Notation}: $\left| x \right|$, ${\left\| {\bf{x}} \right\|_2}$ and  $\left\| {\bf{X}}\right\|_F^2$ denote the absolution value of a scalar $x$, Euclidean norm of a column vector ${\bf{x}} $ and Frobenius norm of matrix ${\bf{X}}$. Tr(${\bf{X}}$), ${{\bf{X}}^T}$, ${{\bf{X}}^H}$, and rank(${\bf{X}}$) denote the trace, transpose, conjugate transpose and rank of the matrix ${\bf{X}}$, respectively. ${\left[ {\bf{x}} \right]_m}$ is the $(m)$-th element of vector ${\bf{x}}$. Diag(${\bf{x}}$) is a diagonal matrix with the entries of ${\bf{x}}$ on its main diagonal. ${\mathbb{C}^{m \times n}}$ denotes the space of ${m \times n}$ complex matrices.

\section{SYSTEM MODEL AND PROBLEM FORMULATION}

\subsection{System Model Description}
We consider a RIS-assisted point-to-point multi-input single-output system, as shown in Fig. 1. Sources $\text{S}_1$ and $\text{S}_2$ are equipped with $N$ transmit antennas and one receive antenna. Both sources operate in FD mode and communicate with each other with the assistance of the RIS. The RIS has $M$ passive reflecting elements. ${\bf{\Phi}}  = {\rm{Diag}}\left\{ {\left[ {{\phi _1}, {\phi _2}, \cdots ,{\phi _M}} \right]} \right\}$ is the phase-shift matrix of the RIS and ${\phi _m} = {e^{j{\theta _m}}}$, where ${\theta _m} \in \left[ {0,2\pi } \right)$ denotes the phase-shift of the $m$-th reflecting element. 

The transmitted signal at each source is 
\begin{equation}
{{\bf{x}}_i} = {{\bf{w}}_i}{x_i},\quad i \in \left\{ \text{1,\,2} \right\},
\end{equation}
where ${x_i}$ denotes the data symbol with normalized power and ${{\bf{w}}_i} \in {\mathbb{C}^{N \times 1}}$ is the beamforming vector at source $\text{S}_i$.

Let ${{\bf{H}}_{{\text{S}_1}\text{R}}} \in {\mathbb{C}^{M \times N}}$, ${{\bf{H}}_{{\text{S}_2}\text{R}}} \in {\mathbb{C}^{M \times N}}$, ${\bf{h}}_{\text{R}{\text{S}_\text{1}}}^H \in {\mathbb{C}^{1 \times M}}$, ${\bf{h}}_{\text{R}{\text{S}_\text{2}}}^H \in {\mathbb{C}^{1 \times M}}$,
${\bf{h}}_{{\text{S}_1}{\text{S}_2}}^H \in {\mathbb{C}^{1 \times N}}$,
${\bf{h}}_{{\text{S}_2}{\text{S}_1}}^H \in {\mathbb{C}^{1 \times N}}$ be the channel from $\text{S}_1$ to RIS, from $\text{S}_2$ to RIS, from RIS to $\text{S}_1$, from RIS to $\text{S}_2$, from $\text{S}_1$ to $\text{S}_2$ and from  $\text{S}_2$ to $\text{S}_1$, respectively\footnote{In this letter, we assume the channel state information (CSI) of all the communication links is perfectly known by each source (CSI estimation methods have been proposed in \cite{ChaEst1,ChaEst2})}. Let ${\bf{h}}_{{\text{S}_1}{\text{S}_1}}^H \in {\mathbb{C}^{1 \times N}}$ and ${\bf{h}}_{{\text{S}_2}{\text{S}_2}}^H \in {\mathbb{C}^{1 \times N}}$ be the SI channels at sources induced by the FD mode. The reflecting SI signal from the RIS can be reasonably neglected or eliminated according to \cite{neg1}, \cite{neg2}, thus it is ignored in the following formulations. 

Then, the received signals at the sources can be expressed as 
\begin{equation}
{y_\text{1}} = ({\bf{h}}_{{\text{S}_2}{\text{S}_1}}^H + {\bf{h}}_{\text{R}{\text{S}_\text{1}}}^H{\bf{\Phi }}{{\bf{H}}_{{\text{S}_2}\text{R}}}){{\bf{x}}_\text{2}} + {\bf{h}}_{{\text{S}_1}{\text{S}_1}}^H{{\bf{x}}_\text{1}} + {z_\text{1}},
\end{equation}
and 
\begin{equation}
{y_\text{2}} = ({\bf{h}}_{{\text{S}_1}{\text{S}_2}}^H + {\bf{h}}_{\text{R}{\text{S}_\text{2}}}^H{\bf{\Phi }}{{\bf{H}}_{{\text{S}_1}\text{R}}}){{\bf{x}}_\text{1}} + {\bf{h}}_{{\text{S}_2}{\text{S}_2}}^H{{\bf{x}}_\text{2}} + {z_\text{2}},
\end{equation}
where ${z_i} \sim \mathcal{C}\mathcal{N} (0,\sigma _i^2),\;i \in \left\{ \text{1,\,2} \right\}$ is the additive white Gaussian noise with zero mean and variance of $\sigma _i^2$ at source $\text{S}_i$. Thus, we can express the achievable rate of the sources in bits second per Hertz (bps/Hz) as follows
\begin{equation}
{R_\text{1}} = {\log _2}\left[ {{\rm{1 + }}\frac{{{{\left| {({\bf{h}}_{{\text{S}_2}{\text{S}_1}}^H + {\bf{h}}_{\text{R}{\text{S}_\text{1}}}^H{\bf{\Phi }}{{\bf{H}}_{{\text{S}_2}\text{R}}}){{\bf{w}}_\text{2}}} \right|}^2}}}{{{{\left| {{\bf{h}}_{{\text{S}_1}{\text{S}_1}}^H{{\bf{w}}_\text{1}}} \right|}^2} + \sigma _\text{1}^2}}} \right].
\end{equation}
\begin{equation}
{R_\text{2}} = {\log _2}\left[ {1 + \frac{{{{\left| { ({\bf{h}}_{{\text{S}_1}{\text{S}_2}}^H + {\bf{h}}_{\text{R}{\text{S}_\text{2}}}^H{\bf{\Phi }}{{\bf{H}}_{{\text{S}_1}\text{R}}}){{\bf{w}}_\text{1}}} \right|}^2}}}{{{{\left| {{\bf{h}}_{{\text{S}_2}{\text{S}_2}}^H{{\bf{w}}_\text{2}}} \right|}^2} + \sigma _\text{2}^2}}} \right].
\end{equation}

The energy consumption of the RIS-assisted HD system can be classified into three major parts: 1) the transmit power; 2) the RIS power consumption; and 3) other hardware static power. However, in order to exploit the advantage of FD mode, the SI needs to be suppressed. Thus the sources would consume extra energy for SIC in FD mode compared to HD mode.

According to \cite{SIC}, analog SIC technology can substract the processed SI from the locally received signals after the propagation-domain SIC. Thus, we can model the SIC power consumption as a linear function of the transmit power as 
\begin{equation}
{P_\text{SIC}} = \xi \left( {{{\left\| {{{\bf{w}}_\text{1}}} \right\|}^2} + {{\left\| {{{\bf{w}}_\text{2}}} \right\|}^2}} \right) + {P_\text{CC}},
\end{equation}
where $\xi $ is the isolation factor (IF) representing the suppression ability of the SIC,  and ${P_\text{CC}}$ is the static power consumption introduced by the SIC circuits \cite{SIC}.

Therefore, we can derive the total energy consumption as 
\begin{equation}
{P_\text{tot}} = \left( {1 + \xi } \right)\left( {{{\left\| {{{\bf{w}}_\text{1}}} \right\|}^2} + {{\left\| {{{\bf{w}}_\text{2}}} \right\|}^2}} \right) + {P_\text{CC}} + M{P_\text{s}} + {P_0},
\end{equation}
where ${P_\text{s}}$ is the hardware-dissipated power at each reflecting element and ${P_0}$ is other static power consumption of the system.

We then define the ratio between the achievable sum rate and the total power consumption as the energy efficiency (EE), which can be expressed as

\begin{equation}\label{EE}
\eta  = \frac{R}{{{P_\text{tot}}}},
\end{equation}
where $R = {R_\text{1}} + {R_\text{2}}$ is the achievable sum rate of the system.
\subsection{Problem Formulation}
In this letter, we aim to maximize the EE by jointly optimizing the active beamforming at the sources and the passive beamforming at the RIS, subject to the minimum data rate demands and maximum transmit power constraints. Mathematically, the optimization problem is formulated as
\begin{subequations}\label{opt1}
	\begin{alignat}{2} 
	\mathcal{P}1:\quad& {\mathop {\max}\limits_{\bm{\Phi},{{{\bf{w}}_\text{1}}},{{{\bf{w}}_\text{2}}}}} &\ &{\eta} \label{opt1A} \\
	& \quad{\textrm {s.t.}}
	&&{R_{i}} \ge {\Gamma _i}, \forall i \in \left\{ \text{1,\,2} \right\},\label{opt1B}\\
	&&&{\left\| {{\bf{w}}_{i}} \right\|^2} \leqslant {P_i^{\max} }, \forall i \in \left\{ \text{1,\,2}\right\}, \label{opt1C}\\
	&&&\left| {{\phi _m}} \right| = 1,\;{\rm{ }}m = 1, \cdots ,M, \label{opt1D}
	\end{alignat}
\end{subequations}	
where ${\Gamma _i}$ is the minimum data rate demand and ${P_i^{\max }}$ is the maximum transmit power at the source  $\text{S}_i$.

The problem $\mathcal{P}1$ is difficult to solve due to the non-convex  constraints in (\text{\ref{opt1B}}) and unit-modulus constraint in (\text{\ref{opt1D}}). In addition, the active and passive beamforming vectors are highly-coupled, which also makes the optimization intractable.

\section{ENERGY EFFICIENCY MAXIMIZATION ALGORITHM DESIGN}
In this section, we adopt the AO method to address the coupling of the active and passive beamforming vectors in problem $\mathcal{P}1$, and decouple the problem $\mathcal{P}1$ into the active beamforming and passive beamforming design subproblems.

\subsection{Active Beamforming Design With Given ${\bf{\Phi }}$}
We define  ${{\bf{F}}_\text{1}} = {({\bf{h}}_{{\text{S}_2}{\text{S}_1}}^H + {\bf{h}}_{\text{R}{\text{S}_\text{1}}}^H{\bf{\Phi }}{{\bf{H}}_{{\text{S}_2}\text{R}}})^H}({\bf{h}}_{{\text{S}_2}{\text{S}_1}}^H + {\bf{h}}_{\text{R}{\text{S}_\text{1}}}^H{\bf{\Phi }}{{\bf{H}}_{{\text{S}_2}\text{R}}})$, ${{\bf{F}}_\text{2}} = {({\bf{h}}_{{\text{S}_1}{\text{S}_2}}^H + {\bf{h}}_{\text{R}{\text{S}_\text{2}}}^H{\bf{\Phi }}{{\bf{H}}_{{\text{S}_1}\text{R}}})^H}({\bf{h}}_{{\text{S}_1}{\text{S}_2}}^H + {\bf{h}}_{\text{R}{\text{S}_\text{2}}}^H{\bf{\Phi }}{{\bf{H}}_{{\text{S}_1}\text{R}}})$, ${{\bf{F}}_\text{11}} = {{\bf{h}}_{{\text{S}_1}{\text{S}_1}}}{\bf{h}}_{{\text{S}_1}{\text{S}_1}}^H$, ${{\bf{F}}_\text{22}} = {{\bf{h}}_{{\text{S}_2}{\text{S}_2}}}{\bf{h}}_{{\text{S}_2}{\text{S}_2}}^H $, ${{\bf{W}}_\text{1}} = {{\bf{w}}_\text{1}}{\bf{w}}_\text{1}^H$ and ${{\bf{W}}_\text{2}} = {{\bf{w}}_\text{2}}{\bf{w}}_\text{2}^H$.

With a given $\bf{\Phi}$, the original problem $\mathcal{P}1$ can be transformed into active beamforming design subproblem as follows
\begin{subequations}\label{opt2}
	\begin{alignat}{2} 
	\mathcal{P}2:\;& {\mathop {\max}\limits_{{{{\bf{W}}_\text{1}}},{{{\bf{W}}_\text{2}}}}} &\ & \frac{{{R}({{\bf{W}}_\text{1}},{{\bf{W}}_\text{2}})}}{{{P_\text{tot}}({{\bf{W}}_\text{1}},{{\bf{W}}_\text{2}})}} \label{opt2A}\\
	&{\textrm {s.t.}}
	&&{\rm{Tr}}({{\bf{F}}_\text{1}}{{\bf{W}}_\text{2}}) \ge {{\Gamma '}_1}({\rm{Tr}}({{\bf{F}}_\text{11}}{{\bf{W}}_\text{1}}) + \sigma _\text{1}^2),\label{opt2B}\\
	&&&{\rm{Tr}}({{\bf{F}}_\text{2}}{{\bf{W}}_\text{1}}) \ge {{\Gamma '}_2}({\rm{Tr}}({{\bf{F}}_\text{22}}{{\bf{W}}_\text{2}}) + \sigma _\text{2}^2),\label{opt2C}\\
	&&&{\rm{Tr}}({{\bf{W}}_\text{i}}) \leqslant {P_i^{\max} }, \forall i \in \left\{ \text{1,\,2}\right\}, \label{opt2D}\\
	&&&{\rm{rank(}}{{\bf{W}}_i})=1, \forall i \in \left\{ \text{1,\,2} \right\}, \label{opt2E}
	\end{alignat}
\end{subequations}
where ${{\Gamma '}_1} = {2^{{{\Gamma}_1}}} - 1$ and ${{\Gamma '}_2} = {2^{{{\Gamma}_2}}} - 1$.

$\mathcal{P}2$ is difficult to solve because the object function is fractional and non-concave. Based on Dinkelbach's method \cite{Dinkerbach}, we introduce an auxiliary variable $\upalpha $ and $\mathcal{P}2$ can be transformed into
\begin{subequations}\label{opt3}
	\begin{alignat}{2} 
	\mathcal{P}2^{\prime}:\;& {\mathop {\max}\limits_{{{{\bf{W}}_\text{1}}},{{{\bf{W}}_\text{2}}}}} &\ & {{R}({{\bf{W}}_\text{1}},{{\bf{W}}_\text{2}})} - \upalpha {{{P_\text{tot}}({{\bf{W}}_\text{1}},{{\bf{W}}_\text{2}})}} \label{opt3A}\\
	&{\textrm {s.t.}}
	&&(\text{\ref{opt2B}})-(\text{\ref{opt2E}}), \label{opt3B}
	\end{alignat}
\end{subequations}
where $\upalpha$ is updated iteratively by
\begin{equation}\label{alpha}
{\upalpha ^{\left( t \right)}} = \frac{{R({\bf{W}}_\text{1}^{\left( t \right)},{\bf{W}}_{\text{2}}^{\left( t \right)})}}{{{P_\text{tot}}({\bf{W}}_\text{1}^{\left( t \right)},{\bf{W}}_{\text{2}}^{\left( t \right)})}}.
\end{equation}

However, the objective function (\text{\ref{opt3A}}) is still non-concave due to ${{{R}({{\bf{W}}_\text{1}},{{\bf{W}}_\text{2}})}}$. However,  ${{{R}({{\bf{W}}_\text{1}},{{\bf{W}}_\text{2}})}}$ can be transformed as the of difference of concave functions (DC) as
\begin{equation}
R({{\bf{W}}_\text{1}},{{\bf{W}}_\text{2}}) = {f_1}\left( {{{\bf{W}}_\text{1}},{{\bf{W}}_\text{2}}} \right) - {f_2}\left( {{{\bf{W}}_\text{1}},{{\bf{W}}_\text{2}}} \right),
\end{equation}
where 
\begin{equation}
\begin{aligned}
{f_1}\left( {{{\bf{W}}_\text{1}},{{\bf{W}}_\text{2}}} \right) &= {\log _2}\left( {{\text{Tr}}\left( {{{\bf{W}}_\text{1}}{{\bf{F}}_\text{11}}} \right) + \sigma _\text{1}^2 + {\text{Tr}}\left( {{{\bf{W}}_\text{2}}{{\bf{F}}_\text{1}}} \right)} \right) \\ & + {\log _2}\left( {{\text{Tr}}\left( {{{\bf{W}}_\text{2}}{{\bf{F}}_\text{22}}} \right) + \sigma _\text{2}^2 + {\text{Tr}}\left( {{{\bf{W}}_\text{1}}{{\bf{F}}_\text{2}}} \right)} \right),
\end{aligned}
\end{equation}
\begin{equation}
\begin{aligned}
{f_2}\left( {{{\bf{W}}_\text{1}},{{\bf{W}}_\text{2}}} \right) &= {\log _2}\left( {{\text{Tr}}\left( {{{\bf{W}}_\text{1}}{{\bf{F}}_\text{11}}} \right) + \sigma _\text{1}^2} \right) \\ & + {\log _2}\left( {{\text{Tr}}\left( {{{\bf{W}}_\text{2}}{{\bf{F}}_\text{22}}} \right) + \sigma _\text{2}^2} \right).
\end{aligned}
\end{equation}

Since ${f_2}\left( {{{\bf{W}}_\text{1}},{{\bf{W}}_\text{2}}} \right)$ is a differentiable concave function, we have its upperbound  at $t$-th iteration as ${f_2}^\prime \left( {{{\bf{W}}_\text{1}},{{\bf{W}}_\text{2}}} \right|{\bf{W}}_\text{1}^{\left( t \right)},{\bf{W}}_\text{2}^{\left( t \right)})$ shown at the top of the page.

\begin{figure*}[ht]
	\small
\begin{equation}\label{upbound}
{f_2}^\prime \left( {{{\bf{W}}_\text{1}},{{\bf{W}}_\text{2}}} \right|{\bf{W}}_\text{1}^{\left( t \right)},{\bf{W}}_\text{2}^{\left( t \right)})  =  {\log _2}\left( {{\text{Tr}}\left( {{\bf{W}}_\text{1}^{\left( t \right)}{{\bf{F}}_\text{11}}} \right) + \sigma _1^2} \right) + \frac{{{\text{Tr}}\left[ {\left( {{{\bf{W}}_\text{1}} - {\bf{W}}_\text{1}^{\left( t \right)}} \right){{\bf{F}}_\text{11}}} \right]}}{{\text{ln2}\left[ {{\text{Tr}}\left( {{\bf{W}}_\text{1}^{\left( t \right)}{{\bf{F}}_\text{11}}} \right) + \sigma _1^2} \right]}}
+ {\log _2}\left( {{\text{Tr}}\left( {{\bf{W}}_\text{2}^{\left( t \right)}{{\bf{F}}_\text{22}}} \right) + \sigma _2^2} \right) + \frac{{{\text{Tr}}\left[ {\left( {{{\bf{W}}_\text{2}} - {\bf{W}}_\text{2}^{\left( t \right)}} \right){{\bf{F}}_\text{22}}} \right]}}{{\text{ln2}\left[ {{\text{Tr}}\left( {{\bf{W}}_\text{2}^{\left( t \right)}{{\bf{F}}_\text{22}}} \right) + \sigma _2^2} \right]}}.
\end{equation}
\hrulefill
\end{figure*}

Therefore, at ($t+1$)-th iteration, we aim to solve the lowerbound maximization problem of $\mathcal{P}2^{\prime}$, which can be expressed as follows:
\begin{subequations}\label{opt4}
	\begin{alignat}{2} 
	\mathcal{P}3:\;& {\mathop {\max}\limits_{{{{\bf{W}}_\text{1}}},{{{\bf{W}}_\text{2}}}}} &\ & {f_1}\left( {{{\bf{W}}_\text{1}},{{\bf{W}}_\text{2}}} \right) - {f_2}^\prime \left( {{{\bf{W}}_\text{1}},{{\bf{W}}_\text{2}}} \right|{\bf{W}}_\text{1}^{\left( t \right)},{\bf{W}}_\text{2}^{\left( t \right)}) \nonumber \\ 
	&&&- \upalpha {{{P_\text{tot}}({{\bf{W}}_\text{1}},{{\bf{W}}_\text{2}})}} \label{opt4A}\\
	&{\textrm {s.t.}}
	&&(\text{\ref{opt2B}})-(\text{\ref{opt2E}}). \label{opt4B}
	\end{alignat}
\end{subequations}

By using semidefinite relaxation (SDR) method \cite{neg1}, we drop the rank-one constraint of  $\mathcal{P}3$ and get the optimized ${\bf{W}}_\text{1}^{\left( {t+1} \right)}$ and ${\bf{W}}_\text{2}^{\left( {t+1} \right)}$ by CVX. It should be noted that in order to ensure convergence, we use eigenvalue decomposition or Gaussian randomization to get the ${\bf{w}}_\text{1}^*$ and ${\bf{w}}_\text{2}^* $ at the end of the overall algorithm.

\subsection{Passive Beamforming Optimization With Given ${{\bf{W}}_{\rm{1}}}$,  ${{\bf{W}}_{\rm{2}}}$}
With given ${{\bf{W}}_{\rm{1}}}$ and ${{\bf{W}}_{\rm{2}}}$, the passive beamforming design subproblem is a sum rate maximization problem, which can be written as follows
\begin{subequations}\label{opt5}
	\begin{alignat}{2} 
	\mathcal{P}4:\;& {\mathop {\max}\limits_{\rm\bm{\theta}}} &\ & {{R}({\rm\bm{\theta}})} \label{opt5A}\\
	&{\textrm {s.t.}}
	&&{{\bm{\theta }}_m} \in \left[ {0,2\pi } \right),\;{\rm{ }}m = 1, \cdots ,M,\label{opt5B}\\
	&&&(\text{\ref{opt1B}}), \label{opt5C}
	\end{alignat}
\end{subequations}
where ${{\bm{\theta }}} = {[{\theta _1}, \cdots ,{\theta _M}]^T}$ is the phase-shifts of the RIS reflecting elements. By denoting ${\bf{q}} = {[{\phi _1}, \cdots ,{\phi _M},1]^H}$, ${{\bf{G }}_\text{1}} = \left[ \begin{array}{l}
{\rm{Diag}}({\bf{h}}_{\text{R}{\text{S}_1}}^H){{\bf{H}}_{{\text{S}_2}\text{R}}}\\ {\bf{h}}_{{\text{S}_2}{\text{S}_1}}^H \end{array} \right]$ and ${{\bf{G }}_\text{2}} = \left[ \begin{array}{l} {\rm{Diag}}({\bf{h}}_{\text{R}{\text{S}_2}}^H){{\bf{H}}_{{\text{S}_1}\text{R}}}\\ {\bf{h}}_{{\text{S}_1}{\text{S}_2}}^H \end{array} \right]$, problem $\mathcal{P}4$ can be transformed as follows
\begin{subequations}\label{opt6}
	\begin{alignat}{2} 
	\mathcal{P}4^{\prime}:\;& {\mathop {\max}\limits_{\bf{Q}}} &\ & {{R}(\bf{Q})} \label{opt6A}\\
	& \;{\textrm {s.t.}}
	&& {\rm{Tr}}({\bf{Q}}{{\bm{\Upsilon }}_\text{1}}) \ge {{\Gamma '}_1}({\rm{Tr}}({{\bf{F}}_\text{11}}{{\bf{W}}_\text{1}}) + \sigma _\text{1}^2),\label{opt6B}\\
	&&&{\rm{Tr}}({\bf{Q}}{{\bm{\Upsilon }}_\text{2}}) \ge {{\Gamma '}_2}({\rm{Tr}}({{\bf{F}}_\text{22}}{{\bf{W}}_\text{2}}) + \sigma _\text{2}^2),\label{opt6C}\\
	&&& {{\left[\bf{Q}\right]}_{m,m}} = 1,\;{\rm{ }}m = 1, \cdots ,M + 1 ,\label{opt6D}\\
	&&&{{\bf{Q}}} \succeq {\bf{0}},\label{opt6E}\\
	&&&{\rm{rank(}}{{\bf{Q}}})=1, \label{opt6F}
	\end{alignat}
\end{subequations}
where ${\bf{Q}} = {\bf{q}}{{\bf{q}}^H}$, ${{\bm{\Upsilon }}_\text{1}} = {{\bf{G }}_\text{1}}{{\bf{W}}_\text{2}}{\bf{G }}_\text{1}^H$, ${{\bm{\Upsilon }}_\text{2}} = {{\bf{G }}_\text{2}}{{\bf{W}}_\text{1}}{\bf{G }}_\text{2}^H$. The constraints (\text{\ref{opt6D}})-(\text{\ref{opt6F}}) are imposed to guarantee that ${\bf{Q}} = {\bf{q}}{{\bf{q}}^H}$ holds after solving problem $ \mathcal{P}4^{\prime}$. 

Problem $ \mathcal{P}4^{\prime} $ is non-convex because of the non-convex rank-one constraint in (\text{\ref{opt6F}}). Hence, we exploit penality method and SCA algorithm to find an optimal rank-one solution.

The non-convex rank-one constraint (\text{\ref{opt6F}}) can be equivalently written as the following equality constraint \cite{Rank}:
\begin{equation}\label{rank}
{\left\| \bf{Q} \right\|_*} - {\bf{\upmu _1 }}({\bf{Q}}) = 0,
\end{equation}
where ${\left\| \bf{Q} \right\|_*} = \sum\limits_l {{\upmu _l}\left( \bf{Q} \right)} $ denote the nuclear norm of $\bf{Q}$ and ${{\upmu _l}\left( \bf{Q} \right)}$ denotes the $l$-th largest eigenvalue.

Note that the equation (\text{\ref{rank}}) indicates that when ${\left\| \bf{Q} \right\|_*} - {\upmu_1}({\bf{Q}}) \to 0$, ${{\bf{Q}}}$ has only one non-zero eigenvalue, and then the rank-one constraint (\text{\ref{opt6F}}) is satisfied. Hence, we transform problem $ \mathcal{P}4^{\prime} $ as follows
\begin{subequations}\label{opt7}
	\begin{alignat}{2} 
	\mathcal{P}5:\;& {\mathop {\max}\limits_{{{{\bf{Q}}}}}} &\ & {{R}(\bf{Q})} - \frac{1}{\beta }[{\left\| \bf{Q} \right\|_*} - {\upmu_1 }({\bf{Q}})] \label{opt7A}\\
	&{\textrm {s.t.}}
	&&(\text{\ref{opt6B}})-(\text{\ref{opt6E}}).\label{opt7B}
	\end{alignat}
\end{subequations}
where $\beta  > 0$ denotes the penalty coefficient used for penalizing the rank-one constraint (\text{\ref{opt6F}}).

The penalty term is in the form of DC functions. For a given point ${{\bf{Q}}^{(k)}}$ in the $k$-th iteration of the SCA method, using first-order Taylor expansion, a convex upper bound for the penalty term can be given by:
\begin{equation}
{\left\| \bf{Q} \right\|_*} - {\upmu_1 }({\bf{Q}})  \leqslant  {\left\| \bf{Q} \right\|_*} - {{\bar \upmu }_1}\left( \bf{Q} \right),
\end{equation}
where ${\bar \upmu_1}({\bf{Q}}) = {\upmu_1}({{\bf{Q}}^{(k)}}) + {\rm{Tr[}}{\mathop{\rm Re}\nolimits} \{ {\bm{\sigma_1 }}({{\bf{Q}}^{(k)}}){\bm{\sigma_1 }}{({{\bf{Q}}^{(k)}})^H}{\rm{(}}{\bf{Q}} - {{\bf{Q}}^{(k)}}{\rm{)}}\} {\rm{]}}$ and ${\bm{\sigma_1 }}({{\bf{Q}}^{(k)}})$ is the corresponding eigenvector of the largest eigenvalue.


Then, the solution at the $(k+1)$-th iteration $ {{\bf{Q}}^{(k + 1)}} $ can be obtained by solving the following problem:
\begin{subequations}\label{opt8}
	\begin{alignat}{2} 
	\mathcal{P}5^{\prime}:\;& {\mathop {\max}\limits_{{{{\bf{Q}}}}}} &\ & {{R}(\bf{Q})} - \frac{1}{\beta }({\left\| \bf{Q} \right\|_*} - {{\bar \upmu }_1}({\bf{Q}})) \label{opt8A}\\
	&{\textrm {s.t.}}
	&&(\text{\ref{opt6B}})-(\text{\ref{opt6E}}).\label{opt8B}
	\end{alignat}
\end{subequations}

In this way, we transform the non-trival problem $ \mathcal{P}4$ into a standard SDP Problem $ \mathcal{P}5^{\prime} $, which can be solved by CVX. 

In summary, we exploit penality method and SCA algorithm to solve the passive beamforming problem, which contains two loops implementation: 1) In the outer loop, we impose a scaling constant $c$ to gradually decrease the penalty coefficient $c{\beta }$ such that ${\left\| \bf{Q} \right\|_*} - {\bf{\upmu_1}}({{\bf{Q}}})$ is forced to zero eventually; 2) In the inner loop, we adopt SCA method to solve $ \mathcal{P}4^{\prime} $ until convergence.  Let ${\varepsilon _1}$ and ${\varepsilon _2}$ be small thresholds. $K$ represents the maximum numbers of iterations. The design of passive beamforming can be shown in Algorithm 1.

\begin{algorithm}[h]
	\caption{Passive Beamforming Algorithm for $ \mathcal{P}4$}
	\begin{algorithmic}[1]
		\State {\textbf{Initialization:}} Get initial $ ({\bf{W}}_\text{1},{\bf{W}}_\text{2}) $ and ${{\bf{Q}}^{(0)}}$. Set $k'=0$ and ${{\beta }^{(0)}} > 0$. 
		\Repeat :\textbf{outer loop}
		\State Set $k=0$;
		\Repeat :\textbf{inner loop}
		\State Solve problem $\mathcal{P}5^{\prime}$ to get ${{\bf{Q}}^{(k+1)}}$ with given $ ({\bf{W}}_\text{1},{\bf{W}}_\text{2}) $ and ${{\beta }^{(k')}}$;
		\State Update $k=k+1$;
		\Until $\left\| {\bf{Q}}^{(k+1)} - {\bf{Q}}^{(k)}\right\|_F^2 < {\varepsilon _1} $ or $k > K$.
		\State Update ${\beta ^{\left( {k' + 1} \right)}} = c{\beta ^{\left( k' \right)}}$;
		\State Update $k'=k'+1$;
		\Until ${\left\|{{\bf{Q}}^{(k'+1)}} \right\|_*} - {\bf{\upmu _1}}({{\bf{Q}}^{(k'+1)}}) < {\varepsilon _2}$.
		\State \textbf{Output:} ${{\bf{Q}}^*}$.
	\end{algorithmic}
\end{algorithm}

Finally, we can get the optimal ${{\bf{q}}^*}$ by eigenvalue decomposition. The phase-shifts of the RIS can be given by
\begin{equation}\label{phase}
{{\bm{\theta}}_m} = -{\text{arg}({\bf{q}}_m^*/{\bf{q}}_{M + 1}^*)},{\rm{ }}\;m = 1, \cdots ,M.
\end{equation}

\subsection{Overall Algorithm}
Our proposed algorithm is summarized in Algorithm 1. ${\varepsilon _d}$ is small threshold and $T$ represents the maximum number of overall iterations. In order to ensure convergence, we use eigenvalue decomposition or Gaussian randomization to get the ${\bf{w}}_\text{1}^*$ and ${\bf{w}}_\text{2}^* $ in Step 9 at the end of the overall algorithm.

\begin{proposition}
	Our proposed Algorithm 2 in step 2-8 is monotonically convergent.
\end{proposition}
\begin{IEEEproof}
	Without loss of generality, we suppose that $\eta$ denote the EE and ${{\bf{W}}} = ({\bf{W}}_\text{1},{\bf{W}}_\text{2}) $, thus we have:
	\begin{equation}
	\eta({{\bf{W}}^{(t+1)}},{{\bm{\theta }}^{(t+1)}})\mathop  \ge \limits^{(a)} \eta({{\bf{W}}^{(t + 1)}},{{\bm{\theta }}^{(t)}})\mathop  \ge \limits^{(b)} \eta({{\bf{W}}^{(t)}},{{\bm{\theta }}^{(t)}}),
	\end{equation}
	where (a) holds because as the penalty coefficient decreases to comparatively small,  the rank-one constraint will be satisfied, thus the SCA-based Algorithm 1 will converges to a stationary point of the original problem \cite{PC}. (b) holds due to optimization of problem $ \mathcal{P}3$ and the update criterion (\text{\ref{alpha}}) of $\upalpha$ is non-decreasing according to \cite{Amplitude}, in which $\upalpha$ stands for EE in power optimization subproblem. 
\end{IEEEproof}

 The overall complexity is ${{\cal O}}\left( {{I_1}\left( {{2N^{3.5}} + {I_2}{I_3}{(M+1)^{4.5}}} \right)} \right)$, where ${I_1}$ is the iteration of AO algorithm, ${I_2}$ and ${I_3}$ are the iterations of inner loop and outer loop for solving $ \mathcal{P}4$.

\begin{algorithm}[h]
	\caption{Overall Algorithm for $\mathcal{P}1$}
	\begin{algorithmic}[1]
		\State {\textbf{Initialization:}} Set initial ${{\bm{\theta }}^{(0)}}$, $ ({\bf{W}}_\text{1}^{(0)},{\bf{W}}_\text{2}^{(0)}) $ and $t = 0$. 
		\Repeat
		\State Calculate ${\upalpha}^{(t)} $ based on (\text{\ref{alpha}});
		\State Use SDR method to solve $\mathcal{P}3$ to get $ ({\bf{W}}_\text{1}^{(t+1)},{\bf{W}}_\text{2}^{(t+1)}) $ with given ${\bm{\theta} ^{(t)}}$ and ${\upalpha}^{(t)} $;
		\State Solve problem  $ \mathcal{P}4$ according to \textbf{Algorithm 1};
		\State Update phase-shift ${{\bm{\theta }}^{(t+1)}}$ based on (\text{\ref{phase}});
		\State Update $t=t+1$;
		\Until The fractional increase of (\text{\ref{opt1A}}) is below ${\varepsilon _d}$ or $t > T$.
		\State Use  eigenvalue decomposition or Gaussian randomization to get $ ({\bf{w}}_\text{1}^*,{\bf{w}}_\text{2}^*) $.
		\State Calculate EE based on (\text{\ref{EE}}).
		\State \textbf{Output:} active and passive beamforming vectors.
	\end{algorithmic}
\end{algorithm}
\section{SIMULATION RESULTS}
This section provides simulation results to verify the performance of our proposed algorithm. We set the number of transmit antennas at each source to 4. We assume that the location of sources $\text{S}_1$, source $\text{S}_2$ and RIS is (0m, 0m) and (200m, 0m) and (20m, 0m). We set $\sigma _\text{A}^2 = \sigma _\text{B}^2 =  - 80$dBm. The large-scale fading is modelled by $PL\left( d \right) = P{L_0}{\left( {d/{d_0}} \right)^{ - \varpi }}$,  where $P{L_0} =  - 30$dB is the path loss at the reference distance ${d_0} = 1$m, $d$ is the distance, and $\varpi $ is the path-loss exponent, which is set to 2.5 for RIS related links and 3.5 for the two sources' direct links \cite{neg1}. We adopt the Rician model for all channels, where the Rician factor is 5dB in SI channel \cite{neg1} and 3dB for others \cite{STAR1}. The path loss of SI channel is set to -100dB due to SIC and the IF of SIC is set to 0.1 \cite{SIC}. We assume the data rate requirements ${R_\text{1}}$ and ${R_\text{2}}$ are 1 bps/Hz. The static power consumption of the SIC circuits ${P_{CC}} = 50 \text{mW}$ \cite{SIC}. The hardware-dissipated power at each reflecting element ${P_s} = 10\text{mW}$ \cite{PowerRIS} and static power consumption ${P_0} = 1000 \text{mW}$ at each source \cite{SWIFT}. We set ${{\beta }^{(0)}} = 100$, ${{c }} = 0.3$, ${\varepsilon _1} = {10^{ - 5}}$, ${\varepsilon _2 = {10^{ - 7}}}$, ${\varepsilon _d = {10^{ - 4}}}$.

The proposed EE optimization scheme for the RIS aided FD system, namely RIS-FD-EE, is compared to the following benchmark schemes:

\begin{itemize}
	\item RIS aided FD system for sum rate maximization (RIS-FD-SR): We assume the system aims to maximize the sum rate without considering the power consumption, which is a special case of our proposed algorithm when $\upalpha$ is set to 0 in active beamforming design.
	\item FD system for EE maximization without RIS (NoRIS-FD-EE): We only optimize the active beamforming at sources without the RIS deployment.
	\item RIS aided HD system for EE maximization (RIS-HD-EE): We assume source $\text{S}_1$ and source $\text{S}_2$ transmit and receive signals in equal time slot. Then the active and passive beamforming vectors in different slots are optimized separately.
\end{itemize}

\begin{figure}
	\centering
	\includegraphics[width=2.5in]{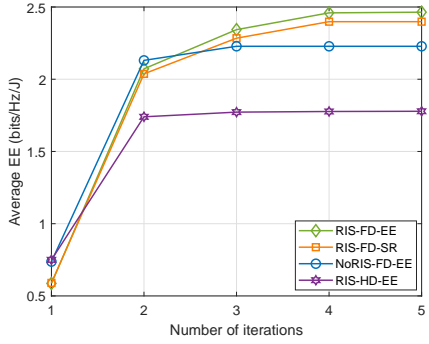}
	\caption{Performance of the proposed method versus the number of iterations.}
\end{figure}

Fig. 2 investigates the convergence of the proposed algorithm, where $M$ is set to 40 and maximal transmit power $P_\text{max}$ is set to 30dBm at both sources. It is shown that all the algorithms increase to convergence. Note that the convergence only needs four-time iterations, which verifies the effectiveness of the proposed algorithm.

\begin{figure}
	\centering
	\includegraphics[width=2.5in]{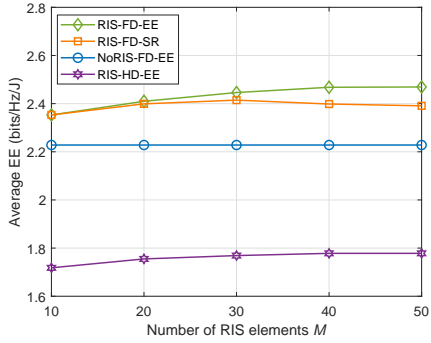}
	\caption{EE versus the number of RIS elements $M$.}
\end{figure}

Fig. 3 shows the EE versus the number of RIS elements, where the maximal transmit power $P_\text{max}$ is set to 30dBm at both sources. The EE of NoRIS-FD-EE scheme is a constant. The other three schemes increase to stable with the number of RIS elements, while our proposed RIS-FD-EE scheme achieves the highest EE.

\begin{figure}[!t]
	\centering
	\includegraphics[width=2.5in]{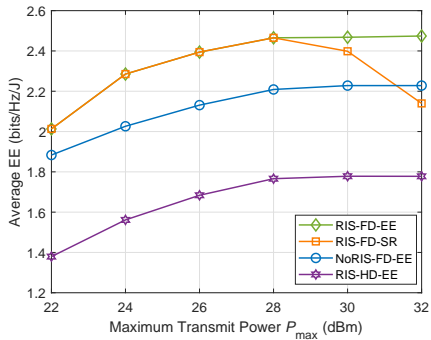}
	\caption{EE versus maximal transmit power $P_\text{max}$.}
\end{figure}

Fig. 4 shows the EE versus different maximal transmit power constraint $P_\text{max}$, which is set equally at both nodes. $M$ is set to 40. It can be seen that the EE of all the schemes increase with $P_\text{max}$ except RIS-FD-SR, which deteriorates at large $P_\text{max}$. That is because the best EE performance does not always require full transmit power.

\section{CONCLUSION}
In this letter, we studied the energy efficiency maximization problem in RIS aided full-duplex system, subject to the minimum data rate demands and maximum transmit power constraints. We divided the optimization problem into active and passive beamforming design subproblems, and adopted the alternative optimization framework to solve them iteratively. We evaluated our scheme under different settings, where extra self-interference cancellation power consumption in full-duplex system was also considered. Simulation results showed the energy efficiency of our scheme outperforms other benchmarks.

\ifCLASSOPTIONcaptionsoff
  \newpage
\fi

\bibliographystyle{IEEEtran}
\bibliography{IEEEabrv,myre}

\end{document}